\documentclass[11pt]{article}

\usepackage{graphicx}
\usepackage{verbatim}
\usepackage{latexsym}
\usepackage{harvard}

\begin{document}

\bibliographystyle{agsm}

\title{{\bf P}={\bf NP}}
\author{Selmer Bringsjord
\& Joshua Taylor\thanks{We're
greatly
indebted to Michael Zenzen for many valuable discussions
about
the {\bf P}=?{\bf NP} problem and digital physics.  Though the two 
arguments
herein establishing {\bf P}={\bf NP} are 
for weal or woe Bringsjord's, Taylor's
astute objections
catalyzed crucial refinements.}\\
Department of Cognitive Science\\
Department of Computer Science\\
The Rensselaer AI \& Reasoning (RAIR) Lab\\
Rensselaer Polytechnic Institute (RPI)\\
Troy NY 12180 USA\\
{\tt \{selmer,tayloj\}@rpi.edu}}
\date{{\footnotesize version of 6.14.04}}
\maketitle
\thispagestyle{empty}

\newpage
\pagenumbering{arabic}
\noindent
The Clay Mathematics Institute offers 
a \$1 million prize for a solution to 
the {\bf P}=?{\bf NP} problem.\footnote{See
{\tt http://www.claymath.org/millennium}.  There are six other
``millennium"
problems; each of these is also associated with a \$1M prize.}
We look forward to receiving our award --- but concede that the
expected format of a solution is an {\em object-level} proof,
not a meta-level argument like what we provide.
On the other hand,
certainly
the winner needn't provide a {\em constructive}
proof that {\bf P}={\bf NP}.\footnote{As
many readers know,
the history of the problem is littered
with failed attempts to provide non-constructive substantiation of
the received view that {\bf P}$\not=${\bf NP}.}
Despite G\"{o}del's recently discovered position
on the matter,\footnote{His position is communicated in a
stunningly prescient
letter he wrote to von Neumann in 1950; this letter is reproduced,
in English, in \cite{sipser.history.pnp.acm}.  G\"{o}del, writing
of course before the modern {\bf P}=?{\bf NP} framework, inquires
as to von Neumann's thoughts about what is today known as the
{\em k}-symbol provability problem.  Let $\phi$ be a formula of
$\cal L$$_I$ (a formula of first-order logic, or just FOL).  We
write $\vdash_k \phi$ provided there is a first-order proof
of $\phi$ of $\le k$ symbols.  G\"{o}del apparently believed 
that it might well be possible to answer questions of the
form ``$\vdash_k \phi$?" in linear or quadratic time.  When
the set here is made explicit and configured so as to allow
for encoding on a Turing machine tape, it's patent that it's
{\bf NP}-complete.  G\"{o}del was quite at home with the idea
that as logic and mathematics progress, machines would increasingly
take over the ``Yes-No" part of the enterprise.  Any notion that
G\"{o}del would have embraced
an argument by analogy from the undecidability of FOL to the
perpetual
intractability of the {\em k}-symbol provability problem
is utterly misguided:  He writes:  ``[I]t would obviously mean
that in spite of the undecidability of the
{\em Entscheidungsproblem}, the mental work of a mathematician
concerning Yes-or-No questions could be completely replaced
by a machine."
}
the general consensus has certainly been that
{\bf P}$\not=${\bf NP}, and many of those brave,
contrarian (and, alas, 
often confused)
souls who have endeavored to show {\bf P}={\bf NP} have sought
to take the beckoning route of exhibiting a polynomial-time
algorithm for one or more of the 1000 or so currently catalogued
{\bf NP}-complete
problems.  This is an exceedingly taxing 
(and, at least hitherto, unproductive) direction to take,
and we eschew it.  We happily concede that constructive success
would have many practical implications, but we are more interested
in the fact of the matter than, say, whether many current
cryptographic schemes can be compromised.  Very well; let's proceed.

In logic and related fields we often speak about
problems in purely abstract terms.
For example, we may declare a problem to be
Turing-solvable, without giving any thought whatsoever
to the {\em embodiment} of a Turing machine able to carry out
a solution.\footnote{An exactly parallel point 
obviously holds of all those
incorporeal
models known to be equivalent to TMs:  register machines,
the $\lambda$-calculus, abaci, etc.}
So we may for instance say that the set $\cal C$ of composite
numbers is Turing-decidable:  that there exists some TM $M$ 
such that, for
every
$n \in$ $\cal N$$ = \{0, 1, 2, \ldots \}$, with
$n$ given to $M$ as input (suitably encoded on
its tape), $M$ produces (say) Y iff $n \in$ $\cal C$,
and N otherwise.  Such facts are routinely confirmed in
the absence of even
a stray, evanescent thought about how $M$ might or might not
be embodied.

However, it's well-known that TMs (and other purely abstract
computers) {\em can} be built.  In fact, one such physical machine
is processing the letters in the present sentence, as I (Selmer) type
them.  We 
may not know for sure that every abstract TM $M^i$ from the 
countably
infinite set of such devices can be physicalized to produce $M^i_p$,
but certainly we {\em do} 
know that 
for every physical TM $M^i_p$ 
able to accomplish some computation,
there exists a corresponding purely mathematical
TM that carries out the same computation (in
the mathematical universe).  This fact will prove
convenient below.

Another well-known fact, one we also find rather helpful, 
is that there are simple
physical processes {\em not reflective of the mathematical 
structure of TMs
and the like}, which nonetheless solve some problems that
are overwhelmingly difficult for TMs and their digital
relatives.  For example, 
the Steiner
Tree problem (STP) is known to be  
{\bf NP}-complete
\citeaffixed{garey.intractability}{see e.g.~pp.~208--209 of}.\footnote{STP
is {\bf NP}-hard when the metric is 
non-discretized, and {\bf NP}-complete
when the metric is discrete.}
Nonetheless, a simple physical process (termed
an {\bf analog computation}\footnote{Analog computers are nothing
new, though they don't get much air time these days.
An elegant example is Claude Shannon's famous differential 
analyzer, which solves
ordinary  differential equations.
A nice discussion of the analyzer can be found in 
\cite{earman}.}) 
can solve it quickly.  STP is the
problem of connecting $n$ points on a plane with a 
graph of minimal
overall length, using junction points if necessary.
The physical process in question runs as follows.
Make two
parallel glass plates, and insert $n$ pins between
the plates to represent the points.  Then dip
the structure into a soap solution, and remove it.
The soap film will connect the $n$ pins in the
minimum Steiner-tree graph
\cite{soapfilm.iwamura}.
Building the structure and the solution
(and the container for the solution)
requires steps linear in the size of $n$, and dipping
and withdrawing make two steps, so despite the fact that
STP is {\bf NP}-complete, the physical process just
described --- let's call it $A^s$ --- is 
apparently carried out well
within $O(n^k)$, for some
constant $k$.

Before starting to read this short paper, you were
probably positive that {\bf P}$\not=${\bf NP}.  If
you've now heard about it for the first time, does
the soap process change your mind?  We didn't think so.
But please reason further with us.

First, some simple notation.  Let's refer to the physical 
version of the STP problem as
$B(STP)$, and the abstract version
as $STP$.  In addition, following usage above, we use
$M$ with or without superscripts to refer to Turing machines,
and $M_p$ to refer to physicalized TMs.  We refer
to analog processes with variable $A$.  Now here is a 
naive proof, functioning as precursor to the more sophisticated
successor given later,
formalizable
in sorted\footnote{E.g., 1 is short for
$\exists x (A(x) \wedge M \mbox{ solves} \ldots)$.}
first-order logic (FOL), that {\bf P}={\bf NP} (where the predicate
letter $N$ is explained later):

\begin{center}
  \begin{small}
\centerline{{\bf The Preliminary Proof}}
    \bigskip
    \begin{tabular}{l|l|l}
      1 & $\exists M$ ($M$ solves STP in polynomial time)
      $\rightarrow$ 
      {\bf P}={\bf NP} & definition of {\bf NP}-completeness\\
      2 & $\exists A$ ($A$ solves $B(STP)$ in polynomial time $\wedge 
      \, N(A))$ & derivable, e.g., by existential introduction\\
	      & & from soapfilm process, i.e., $A^s$\\
      3 & $\exists A$ ($A$ solves $B(STP)$ in polynomial time $\wedge N(A)$)
      $\rightarrow$ &\\
      & $\exists M_p (M_p $ solves $B(STP)$
      in polynomial time) & digital physics; see below \\
      4 & $\exists M_p (M_p$ solves $B(STP)$
      in polynomial time) $\rightarrow$ &\\
      & $\exists M (M$ solves $STP$
      in polynomial time) & unassailable; see justification 3rd \P\\
      \hline
      5 & {\bf P}={\bf NP} & 1--4 (full FOL derivation trivial)\\
    \end{tabular}
    \bigskip
  \end{small}
\end{center}

There is no question that the reasoning here can be certified as
formally valid (e.g., using an automated proof checker).
The only question is whether the premises are true.  If they are,
the problem is at long
last solved in the affirmative.  Are the premises true?

Line 2, note, isn't a premise, but rather an intermediate 
conclusion; however,
there are two routes to this conclusion.
As noted in the
justification column, one possible inference to line 2 is
from $A^s$, where this constant is replaced
by the variable $A$, and existential introduction
is used (we assume for certification a natural deduction
calculus, with rules for introducing and eliminating
truth-functional connectives and quantifiers; a nice
system of this sort is $\cal F$, from Barwise
\& Etchemendy \citeyear*{lpl}).
The second route
takes account of what we regard to be
self-evident:  Surely $A^s$ is just the tip of the iceberg,
with myriad analog processes out there in our physical 
universe waiting to be discovered and harnessed (though
presumably most will remain undetected for eternity).
This view can be derived from a 
sampling
assumption, according to which a finding like $A^s$ must be
a random sampling from some small proper subset of the
(probably infinite) set of all candidate processes available
in the cosmos.
We don't pursue this derivation herein.  Interested readers should
consult a parallel form of argument explored in theoretical
physics \citeaffixed{antropic.bias.bostrom}{see e.g.}.

We anticipate that some will be uncomfortable with the view
that there exists a process $A$ that accommodates ever
greater values for $n$.  In light of this,
we move now to the more sophisticated
of our two proofs:  First, note that 
full specification of our proof in FOL does
include universal quantification over $\cal N$, and
a corresponding index
for $STP$ and $B(STP)$, as for example in 
what line 1, unpacked, becomes:

$$\exists A \forall n (\ldots B(STP)_n \ldots) \ldots$$

\noindent But to us, this simply calls for a 
natural variant of induction on $\cal N$.
The base clause is trivial.  Given the induction
hypothesis, it's exceedingly hard to see how $A^s$
cannot succeed on $n+1$ if the minimal graph has
been found for $n$ physicalized points.
Whatever underlying principles of physics
generate the graph in the case of $n$ surely can be
employed to generate it for $n+1$.  Even if the physical
laws governing {\em our} universe are such that there is
some point $n+1$ at which $A^s$ fails, surely it's physically
{\em possible} that this failure {\em not} occur.
This implies
that there is a more formidable second 
proof that employs modal logic 
\cite{chellas,hughes.cresswell}.
If we let $\Diamond_p$ refer to physical
possibility in a manner that parallels the straight $\Diamond$
of logical possibility from modal logic,\footnote{We assume
a normal S5 version of the $\Diamond$ operator.}
then the
modal version of line 3 is

\begin{small}
$$\Diamond_p \exists A \forall n \mbox{(}A \mbox{ solves } B(STP)_n 
\mbox{ in polynomial time } \wedge N(A)\mbox{)}
      \rightarrow$$
$$\Diamond_p \exists M_p \forall n \mbox{(}M_p \mbox{ solves } B(STP)_n
      \mbox{ in polynomial time)}$$
\end{small}

\noindent and this technique can be easily propagated
through the original proof to produce the more
circumspect one.  In this
modal proof, line 4 becomes the key principle that if 
it's physically possible that a physical
TM solve $B(STP)_n$
in polynomial time, then there exists (in the mathematical
universe) a TM that solves $STP_n$ in polynomial time.
This principle would appear to be invulnerable.  Summing
up, we have:

\begin{center}
  \begin{footnotesize}
\centerline{{\bf The Modalized Proof}}
    \bigskip
    \begin{tabular}{l|l|l}
      1$'$ & $\exists M \forall n (M$ solves $STP_n$ in polynomial time)
      $\rightarrow$ 
      {\bf P}={\bf NP} & definition of {\bf NP}-completeness\\
      2$'$ & $\Diamond_p \exists A \forall n$ ($A$ solves $B(STP)_n$ in
polynomial time
$\wedge 
      \, N(A))$ & derivable, e.g., by induction and existential \\
	      & & introduction from soapfilm process, i.e., $A^s$\\
      3$'$ & $\Diamond_p \exists A \forall n$ ($A$ solves $B(STP)_n$ in
polynomial time $\wedge N(A)$)
      $\rightarrow$ &\\
      & $\Diamond_p \exists M_p \forall n (M_p $ solves $B(STP)_n$
      in polynomial time) & digital physics; see below\\
      4$'$ & $\Diamond_p \exists M_p \forall n (M_p$ solves $B(STP)_n$
      in polynomial time $\rightarrow$ &\\
      & $\exists M \forall n (M$ solves $STP_n$
      in polynomial time)) & unassailable; see justification 3rd \P\\
      \hline
      5$'$ & {\bf P}={\bf NP} & 1$'$--4$'$ (full FOL derivation trivial)\\
    \end{tabular}
    \bigskip
  \end{footnotesize}
\end{center}

Please note that the modal version of our argument provides
complete immunity from an objection that the physical universe is
finite, and that therefore no analog process can scale up through
all natural numbers as inputs for the minimal graph to
be generated.  The dominant view among theoretical
physicists appears to be that the theory of inflation
\cite{birth.vilenkin,inflation.guth}
holds
(which renders it likely that the universe is infinite).
But we need not take a stand on the issue.
All we need is what follows immediately
from the fact that the theory of inflation, whether or not
true, is certainly coherent:
namely, that it's physically {\em possible} that
the universe is infinite.

By ``digital physics'' for the justification of premise 3/3$'$, we have
in mind the position that the physical universe is fundamentally
a vast physical computer --- or, if you like, a computer composed 
of computers,
which are in turn composed of computers, and so on.
This view has been recently affirmed by \citeasnoun{wolfram.new},
but \citeasnoun{digital.mechanics.fredkin}
advanced the view long ago
(and continues to energetically defend it now), and Feynman
\citeyear{feynman.simphysics} seems to have
embraced the view as well.  Even Einstein can be read
as having affirmed
the digital physics position.  Though premise 3/3$'$ refers to physicalized
Turing machines, most digital models in physics are based on
cellular automata, but this is of no matter:  it's well-known that every
cellular automaton can be recast as a TM.\footnote{The
transformation preserves polynomial-time processing, as cognoscenti
know.  For others, a sketch:
Use an $n$-dimension TM in which $n$ is high enough to sufficiently
represent the CA which is the universe.
Let each cell of the TM's tapes
represent a cell of the CA. The alphabet of the TM 
will contain some representation
of all states for the CA's cells. The computation performed by the CA is
finite, so the TM's states
are as well. It is known that the transformation
from multidimensional Turing machines to standard 
Turing machines is a polynomial
transformation. If the computation performed in each cell of the CA
is in class {\bf P}, the equivalent TM will be in class {\bf P}.
}

Our argument shows that
if {\bf P}$\not=${\bf NP}, digital physics is incorrect.
Since it must be true that all physical phenomena can
in principle be modeled in information-processing terms of {\em some}
kind, 
{\bf P}$\not=${\bf NP}
thus immediately implies,
courtesy of our arguments, that hypercomputational processes exist in
the physical universe.\footnote{Physical phenomena that can
be rigorously modeled only via information processing above
the {\bf Turing Limit} \cite{siegelmann.book,superminds} would
be phenomena
calling for hypercomputational machines.
Just as the
class of mathematical devices equivalent to TMs is infinite, so also
there are an infinite number of hypercomputational machines.  Examples
include {\bf analog chaotic neural nets} \cite{siegelmann.sontag} and
{\bf infinite time Turing machines} \cite{hamkins.lewis.ittms}.
Other examples include {\bf analog ``knob" TMs} \cite{para}
and {\bf accelerated TMs} \cite{copeland.accelerated.tms}.}
If you believe, as many do, that hypercomputational
processes are always merely mathematical, and never physically real,
you can't be rational and at the same time refuse to accept our
case for {\bf P}={\bf NP}.

Perhaps you do indeed refuse to accept the digital physics view, and
have no qualms about physical hypercomputation.
It was with skeptics like you in mind that the predicate 
$N(\,\,\,)$ (for ``normal") was included in
our proofs.  While many are perhaps right to point out, {\em contra} 
Wolfram
and company, that some physical phenomena (e.g., those associated
with quantum mechanics) are so bizarre and complicated that they
resist formalization in TM-level computational models, the fact of 
the matter
is that the analog process we exploit is a painfully simple macroscopic
phenomenon --- as we say, a ``normal''
physical process.  The burden of proof is surely on those who would 
maintain
that the formal machinery of digital physics is insufficient to model
something as straightforward as submerging  nails in,
and retrieving them from, a bucket of soapy water.

\newpage
\bibliography{/personal/BIB/main70}

\end{document}